\documentclass[12pt,reqno]{article}
\usepackage{times}

\usepackage{amsmath}
\usepackage{amsthm}
\usepackage{amsfonts}
\usepackage{amssymb}
\usepackage{epic}
\usepackage{eepic,epsfig}
\usepackage{times, macros}
\usepackage{graphics}

\theoremstyle{plain}

\theoremstyle{remark}

\newcommand{\ra}{\to}
\newcommand{\fr}[2]{{\textstyle \frac{#1}{#2} }}

\newcommand{\fsl}{{\mathfrak s}{\mathfrak l}}

\newcommand{\al}{\alpha}

\newcommand{\ga}{\gamma}

\newcommand{\de}{\delta}
\newcommand{\De}{\Delta}

\newcommand{\la}{\lambda}

\newcommand{\vf}{\varphi}

\newcommand{\pa}{\partial}

\newcommand{\CM}{{\mathcal M}}

\newcommand{\CO}{{\mathcal O}}

\newcommand{\CT}{{\mathcal T}}

\newcommand{\CW}{{\mathcal W}}

\newcommand{\CZ}{{\mathcal Z}}

\newcommand{\SH}{{\mathsf H}}

\newcommand{\SL}{{\mathsf L}}

\newcommand{\ST}{{\mathsf T}}

\newcommand{\sv}{{\mathsf v}}

\newcommand{\BC}{{\mathbb C}}

\newcommand{\rf}[1]{(\ref{#1})}

\begin{document}\thispagestyle{empty}
\title{Classical conformal blocks and isomonodromic deformations}
\author{J\"org Teschner}
\address{Department of Mathematics, \\
University of Hamburg, \\
Bundesstrasse 55,\\
20146 Hamburg, Germany,\\[1ex]
and:\\[1ex]
DESY theory, \\
Notkestrasse 85,\\
20607 Hamburg,
Germany}
\maketitle

\section{Introduction}

The classical limit 
of conformal field theories is of interest for various 
reasons. It gives the link between a Lagrangian description of a CFT and
the abstract representation-theoretic definition of its correlation functions provided
by the bootstrap approach. It is also crucial for understanding several aspects of 
the geometry encoded in the correlation function of conformal field theory. 
This is relevant in particular for various 
models of holographic 
correspondences between two-dimensional CFT and three-dimensional
quantum gravity investigated in the context of the 
$\mathrm{AdS}_3-\mathrm{CFT}_2$-correspondence.

We are here going to demonstrate that
conformal field theory is related to the isomonodromic deformation
problem in the limit $c\ra\infty$. The Schlesinger system describes 
monodromy preserving deformations of $2\times 2$ first order matrix 
differential equations on $\mathbb{P}^1$ with $n$ regular singularities. 
It has an alternative description 
known as the Garnier system 
describing  the isomonodromic deformations of the 
second order ODE  naturally associated to the first order matrix differential equation. 
We refer to \cite{IKSY} for a review and further references.
It will be shown that one may 
describe the leading asymptotics of Virasoro conformal blocks with a suitable number
of insertions of degenerate representations in terms of the generating function 
for a change of coordinates between two natural sets of 
Darboux coordinates for the Garnier system. One set of coordinates is  natural
for the Hamiltonian  formulation of the Garnier system  \cite{IKSY}, the other coordinates 
will be called complex Fenchel-Nielsen
coordinates parameterising  the space of monodromy data 
of the differential equation on $C_{0,n}=\mathbb{P}^1\setminus\{z_1,\dots,z_n\}$. 

The results of this paper characterise the leading classical asymptotics of 
Virasoro conformal blocks completely, and clarify in which sense conformal field theory
represents a quantisation of the isomonodromic deformation problem. 

\section{The Garnier system}

\subsection{Basic definitions}

The Garnier system describes monodromy preserving deformations of the differential equations  
$(\pa^2_y+t(y))\psi(y)=0$ with  $t(y)$ of the form
\begin{equation}\label{t-form}
t(y):=\sum_{r=1}^n
\left(\frac{\de_r}{(y-z_r)^2}-\frac{H_r}{y-z_r}\right)-
\sum_{k=1}^{n-3} \left(\frac{3}{4(y-u_k)^2}-\frac{v_k}{y-u_k}
\right).
\end{equation}
The differential equation $(\pa^2_y+t(y))\psi(y)=0$ has regular 
singular points at $y=z_r$ and $y=u_k$. The parameters $\de_r$ will be fixed once and for all.
The singular points at $y=u_k$ are special.
They are called {\it apparent singularities} if 
the parameters $(u_r,v_r,H_r)$ are not independent but related through the constraints
\begin{align}\label{HJ}
&v_k^2+\check{t}_{k}^{}(u_k)=0,\qquad 
\check{t}_k(u_k):=\lim_{y\ra u_k}\left(t(y)+\frac{3}{4(y-u_k)^2}-\frac{v_k}{y-u_k}
\right),
\end{align}
These constraints imply that the monodromy around $y=y_k$ is $-\mathrm{id}$.
Indeed, having monodromy $-\mathrm{id}$  is easily  seen 
to be equivalent to the fact that there exists a solution $\psi$ of $(\pa^2_y+t(y))\psi(y)=0$ which
has the form $\psi(y)=\exp\big(-\frac{1}{2}\int^y dy'\eta(y')\big)$, 
with $\eta(y)$ of the local form $\eta(y)=\sum_{l=0}^\infty (y-u_k)^{l-1}\eta_l$ 
satisfying the Ricatti equation $t(y)=-\frac{1}{4}\eta^2+\frac{1}{2}\eta'$. The Ricatti equation
determines the  coefficients $\eta_l$ recursively in terms of the expansion coefficients of 
$t(y)=\sum_{l=0}^\infty t_l (y-u_k)^{l-2}$. When $t_0=-\frac{3}{4}$ one finds the relation $v^2+\check{t}_k(u_k)=0$
as necessary and sufficient condition for the existence of a solution to the recursion relations following from the 
Ricatti equation.

In order to define the Garnier system we will choose the number of apparent singularities to be 
$d=n-3$. More general values of $d$ will be discussed later.
We shall furthermore assume that the differential equation $(\pa^2_y+t(y))\chi(y)=0$ is regular
at $y=\infty$, which implies
\begin{align}\label{globalsl2}
&\sum_{r=1}^nz^l_r\big(z_rH_r-(l+1)\de_r\big)-\sum_{k=1}^{n-3}u^l_k\big(u_kv_k-(l+1)\fr{3}{4}\big)=0.
\end{align}
where $l=-1,0,1$.
The constraints \rf{globalsl2} determine three of the $H_r$, in what follows usually chosen to be $H_n$, 
$H_{n-1}$ and $H_{n-2}$ in terms of $H_r$, $r=1,\dots,n-3$.
Equations
\rf{HJ} can then be solved allowing us to express $H_r$ as a function $H_r=H_r(\mathbf{u},\mathbf{v},\mathbf{z})$,
of $\mathbf{u}=(u_1,\dots,u_{n-3})$, $\mathbf{v}=(v_1,\dots,v_{n-3})$ and
$\mathbf{z}=(z_1,\dots,z_{n})$. The 
 ``potential'' $t(y)$ thereby gets determined as a function
$t(y)= t(y|\mathbf{u},\mathbf{v},\mathbf{z})$.

It can be shown \cite{Ok,IKSY} 
that the Hamiltonian equations of motion 
\begin{equation}\label{Hameom}
\frac{\pa u_r}{\pa z_s}=\frac{\pa H_s}{\pa v_r}\,,\qquad
\frac{\pa v_r}{\pa z_s}=-\frac{\pa H_s}{\pa u_r}\,.
\end{equation} 
ensure that the monodromy of the differential 
equation $(\pa^2_y+t(y))\chi(y)=0$ stays constant 
under variations of the parameters $(\mathbf{u},\mathbf{v})$
satisfying \rf{Hameom}.
The  
coordinates $(\mathbf{u},\mathbf{v})$ 
are Darboux coordinates for the 
natural symplectic structure of the Garnier system.

\subsection{Relation to the Schlesinger system}

More widely known than the Garnier system may be the 
Schlesinger system describing isomonodromic deformations
of holomorphic $\fsl_N$-connections  of the
form $\pa_y-A(y)$ on $C_{0,n}=\mathbb{P}^1\setminus\{z_1,\dots,z_n\}$,  
with matrix-valued functions $A(y)$ of the form
\begin{equation}\label{Aform}
A(y)\,=\,\sum_{r=1}^n \frac{A_r}{y-z_r}\,.
\end{equation}
We will assume that  $A_1,\dots A_n$ satisfy $\sum_{k=1}^nA_k=0$. 
Allowing that the  residues $A_r$ depend on the parameters $\mathbf{Z}=(z_1,\dots,z_n)$ 
in a suitable way, one may ensure that the monodromy of $\pa_y-A(y)$ does not depend
on $\mathbf{Z}$.
The Schlesinger system is the system of nonlinear partial differential equations 
describing how to cancel 
variations of $z_r$ by corresponding variations of the residues $A_s$, $s=1,\dots,n$.

The Garnier system is nothing but the Schlesinger system for $N=2$ in disguise. The relation 
between these two dynamical systems is found by representing the holomorphic connection 
$\pa_y-A(y)$ containing the dynamical variables of the Schlesinger system in the form
$g^{-1}(\pa_y-B(y))g$, with 
$B(y)=\big(\begin{smallmatrix} 0 & -t \\ 1 & 0 \end{smallmatrix}\big)$. 
It is straightforward to find a matrix function $g=g(y)$ relating 
$B(y)$ to 
$A(y)=\big(\begin{smallmatrix} A_0 &  A_+\\  A_-& -A_0 \end{smallmatrix}\big)$
in this way, provided one allows $g(y)$ to have square-root branch points at the 
zeros of $A_-(y)$. 
It is furthermore
straightforward to show that the function $t(y)$ representing the only non-constant 
matrix element of $B(y)$ will have a singularity of the form 
\begin{equation}\label{appsing}
t(y)=-\frac{3}{4(y-u)^2}+\frac{v}{y-u}+\check{t}(u)+\CO((u-v)^1),
\end{equation}
near each simple zero $u$ of $A_-(y)$. The coefficients  $v$ and $\check{t}(u)$ appearing in the 
Laurent expansion \rf{appsing} must satisfy the relation $v^2+\check{t}(u)=0$, which is 
the necessary and sufficient condition for the solutions to $(\pa_y^2+t(y))\psi=0$ to have 
monodromy proportional to $-1$ around $y=u$. 
Denoting the $n-3$ zeros $A_{-}(y)$ generically has by $\mathbf{u}=(u_1,\dots,u_{n-3})$
and the corresponding residues by $\mathbf{v}=(v_1,\dots,v_{n-3})$ one 
recovers exactly the form of the function $t(y)$ considered in the theory of 
the Garnier system.

The gauge transformation from $A(y)$ to $B(y)$ described above defines a map
from the Schlesinger
system to the Garnier system. It is known that this map relates the natural symplectic structures \cite{DM}.

\subsection{Complex-Fenchel-Nielsen coordinates for moduli spaces of flat connections}\label{Darboux}

The monodromy data represent the conserved quantities which remain constant in the 
Hamiltonian flows of the Garnier system, by definition. 
The goal of this subsection is to introduce useful coordinates for the space of monodromy 
data.

Holonomy map and Riemann-Hilbert correspondence between flat connections $\pa_y-A(y)$ and
representations $\rho:\pi_1(C_{0,n})\ra {\rm SL}(2,\BC))$  relate
the moduli space $\CM_{\rm flat}(C_{0,n})$ of flat $\fsl_2$-connections on $C_{0,n}$ to the so-called character
variety
$\CM_{\rm char}(C_{0,n})={\rm Hom}(\pi_1(C_{0,n}),{\rm SL}(2,\BC))/{\rm SL}(2,\BC)$.
One set of useful sets of coordinates for
$\CM_{\rm flat}(C_{0,n})$ is given by the trace functions
$L_{\ga}:=\operatorname{\rm tr}\rho(\ga)$ associated to 
simple closed curves $\ga$ on $C_{0,n}$. 

Minimal sets of trace functions that can be used to
parameterise $\CM_{\rm flat}(C_{0,n})$ can be identified using
pants decompositions. A pants decomposition is defined by cutting 
$C_{0,n}$ along $n-3$ simple non-intersecting closed curves $\ga_k$. 
Each curve $\ga_k$ separates two pairs of pants, the union of which 
will be a four-holed sphere $C_{0,4}^k$. As 
$\mathrm{dim}(\CM_{\rm flat}(C_{0,n}))=2(n-3)$ it suffices to introduce two coordinates 
for the flat connections on each $C_{0,4}^k$.

\begin{figure}[h]
\epsfxsize13.5cm
\centerline{\epsfbox{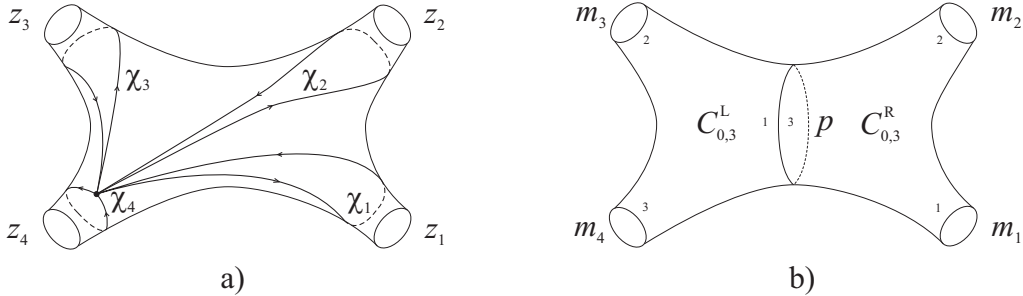}}
\caption{\it Basis of loops of $\pi_1(C_{0,4})$ and the decomposition $C_{0,4}=C_{0,3}^L\cup C_{0,3}^R$.}
\label{c04}\vspace{.3cm}
\end{figure}

Let us therefore restrict attention to the case $n=4$ 
in the following. 
Conjugacy classes of irreducible representations of $\pi_1(C_{0,4})$ are uniquely specified by
seven invariants
\begin{subequations}
\begin{align}\label{Mk}
&L_a=\operatorname{Tr} M_a=2\cos2\pi m_a,\qquad a=1,\ldots,4,\\
&L_s=\operatorname{Tr} M_1 M_2,\qquad L_t=\operatorname{Tr} M_1 M_3,\qquad L_u=\operatorname{Tr} M_2 M_3,
\end{align}
\end{subequations}
generating the algebra of invariant polynomial functions on $\CM_{\rm char}(C_{0,4})$. The monodromies $M_r$ are associated to the
curves $\ga_r$ depicted in Figure \ref{c04}. 
These trace functions satisfy the quartic equation
\begin{align}
 \label{JFR}
& L_1L_2L_3L_4+L_sL_tL_u+L_s^2+L_t^2+L_u^2+L_1^2+L_2^2+L_3^2+L_4^2=\\
 &\nonumber \quad=\left(L_1L_2+L_3L_4\right)L_s+\left(L_1L_3+L_2L_4\right)L_t
+\left(L_2L_3+L_1L_4\right)L_u+4.
\end{align}
The affine algebraic variety defined by (\ref{JFR}) is a concrete representation for
the character variety of $C_{0,4}$. For fixed choices of $m_1,\ldots,m_4$ in \rf{Mk}
equation \rf{JFR} describes the character variety as a cubic surface in $\BC^3$.

This surface
admits a parameterisation in terms of coordinates $(\la,\kappa)$ of the form
\begin{align}\label{classWT}
L_s\,=\,2\cos 2\pi \lambda\,,\qquad
\begin{aligned}
& (\sin (2\pi  \lambda))^2\,L_t\,=\,
C_t^+(\lambda)\,e^{i\kappa}+C_t^0(\lambda)+C_t^-(\lambda)\,e^{-i\kappa}\,,\\
&  (\sin (2\pi  \lambda))^2\,L_u\,=\,
C_u^+(\lambda)\,e^{i\kappa}+C_u^0(\lambda)+C_u^-(\lambda)\,e^{-i\kappa}\,,
\end{aligned}
\end{align}
where
\begin{subequations}\label{Cepdef}
\begin{align}
C_u^{\pm}(\la)&=-4\prod_{s=\pm 1} \sin\pi(\la+s(m_1\mp m_2))\sin\pi(\la+s(m_3\mp m_4))\,,\\
C_u^0(\la)&={2 }\,
\big[\cos 2\pi  m_2 \cos 2\pi m_3 + \cos 2\pi  m_1 \cos 2\pi m_4\big]\\
&\quad- {2 \cos 2\pi  \la}
\big[\cos 2\pi  m_1 \cos 2\pi m_3 + \cos 2\pi  m_2 \cos 2\pi m_4\big]
\, .\notag
\end{align}
\end{subequations}
together with similar formulae for $C_t^{k}$, $k=\pm,0$.

Using pants decompositions as described above one may define trace coordinates $L_{k,s}$, $L_{k,t}$
and $L_{k,u}$ for each four-holed sphere $C_{0,4}^k$ defined above. In this way one may define
a pair of coordinates $(\kappa_k,\la_k)$ associated 
to each cutting curve $\ga_k$, $k=1,\dots,n-3$.
Taken together, the tuples $\mathbf{k}=(\kappa_1,\dots,\kappa_{n-3})$
and $\mathbf{l}=(\lambda_1,\dots,\la_{n-3})$ form a system of coordinates for $\CM_{\rm flat}(C)$.
It is known that the coordinates $(\mathbf{k},\mathbf{l})$ are a set of Darboux coordinates for the 
moduli space $\CM_{\rm flat}(C)$ of flat $\mathrm{SL}(2,\BC)$-connections on $C_{0,n}$ \cite{NRS},
bringing the natural symplectic structure on this space to the simple form
\begin{equation}
\Omega=\frac{1}{4\pi}
\sum_{r=1}^{n-3}d\kappa_r\wedge d\lambda_r\,.
\end{equation}
We note that the trace functions $L_k=2\cos(2\pi\la_k)$ are globally well-defined, and that 
the Hamiltonian flows generated by the functions $\la_k$ are linear in the variables $\kappa_k$.
One may therefore view the coordinates 
$(\mathbf{k},\mathbf{l})$ as action-angle variables making the integrable structure of
the character variety manifest.

\section{Classical limit of Virasoro conformal blocks}\label{sec:Garnier}

We are now going to explain how the Garnier system arises in  the limit
$c\ra \infty$ of certain Virasoro conformal blocks with degenerate field insertions.

\subsection{Conformal blocks with degenerate fields}\label{Set-up}

Conformal blocks, the holomorphic building blocks of physical correlation functions in conformal 
field theories, can be defined as solutions to the conformal Ward identities \cite{BPZ}. 
Our discussion will be brief, referring for the relevant background on conformal field theory 
to reviews such as \cite{T17a}. Conformal blocks can be defined for all punctured Riemann
surfaces, with representations of the Virasoro algebra assigned to each puncture. We will
consider highest weight representations $V_\al$ of the Virasoro algebra with central charge 
$c=1+6Q^2$ with highest weight vector $e_\al$ satisfying $L_{n}e_\al=\de_{n,0}\De_\al e_\al$ for $n\geq 0$,
where $\De_\al=\al(Q-\al)$. It will be convenient to represent the parameter $Q$
as $Q=b+b^{-1}$.

We will  consider Virasoro conformal blocks on $C_{0,n-3+d+2}$ 
with generic representations $V_{\al_r+1/2b}$ assigned to the punctures at $z_r$, $r=1,\dots,d$.
Representation $V_{\al_r}$ are assigned to  the remaining punctures $z_r$, $r=d+1,\dots,n$,
and we will assume that   $z_{n-2}=1$, $z_{n-1}=0$, $z_n=\infty$ to simplify some formulae.
Degenerate representations  $V_{-1/2b}$ are associated to $d$ punctures at $u_k$, $k=1,\dots,d$,
and degenerate representations $V_{-b/2}$ are associated to the punctures at $y$ and $z_0$, 
respectively. The corresponding chiral partition functions satisfy the following differential equations
\begin{subequations}\label{BPZ+}
\begin{align}
&\bigg(\frac{1}{b^2}\frac{\pa^2}{\pa y^2}+\ST(y)\bigg)\CZ(\mathbf{u},\hat{\mathbf{z}};y)=0,
\label{BPZlight}\\
&\qquad\ST(y):=\sum_{r=0}^n
\left(\frac{\De_r}{(y-z_r)^2}+\frac{1}{y-z_r}\frac{\pa}{\pa z_r}\right)-
\sum_{k=1}^{d}\left(\frac{3b^{-2}+2}{4(y-u_k)^2}-\frac{1}{y-u_k}
\frac{\pa}{\pa u_{k}}\right),\notag \\
& \bigg( b^2\frac{\pa^2}{\pa u^2_k}+\check{\ST}_k(u_k) -
\frac{3b^2+2}{4(u_k-y)^2}+\frac{1}{u_k-y}
\frac{\pa}{\pa y}\bigg)\CZ(\mathbf{u},\hat{\mathbf{z}};y)=0,\quad k=1,\dots,d,\label{BPZheavy}\\
&\qquad\check{\ST}_k(u_k):=\lim_{y\ra u_k}\left(\ST(y)+\frac{3b^{-2}+2}{4(y-u_k)^2}-\frac{1}{y-u_k}
\frac{\pa}{\pa u_{k}}
\right),\notag \\
&\sum_{r=0}^nz^l\bigg(z_r\frac{\pa}{\pa {z_r}}+(l+1)\De_r\bigg)+
\sum_{k=1}^{n-3}u_k^l\bigg(u_k\frac{\pa}{\pa {u_k}}+(l+1)\De_{-1/2b}\bigg)\notag\\&
\qquad\qquad
+y^l\bigg(y\frac{\pa}{\pa {y}}+(l+1)\De_{-b/2}\bigg)
=0,\qquad l=-1,0,1,
\label{projective}\end{align}
\end{subequations}
using tuple notations $\hat{\mathbf{z}}=(z_0.z_1,\dots,z_{n-3})$, $\mathbf{u}=(u_1,\dots,u_d)$.
Equation \rf{BPZlight} reflects the decoupling of the null-vector in the Verma-module associated to the 
representation $V_{-b/2}$ assigned to the point $y$, while \rf{BPZheavy} is equivalent to the decoupling
of the null-vectors in the representations $V_{-1/2b}$ associated to the   
punctures $u_k$, $k=1,\dots,d$. Equation \rf{projective} simply reflects the global $\mathrm{SL}(2)$-invariance
on the sphere.
It turns out that the insertion of the representations $V_{-b/2}$ modifies 
the conformal blocks only mildly in the limit $b\ra 0$. We will use the representation at $y$ as a ``probe'', 
exploiting the information provided by the associated differential equation, and the analytic
continuation of its solutions. The  representation at $z_0$ will only serve the task to 
define a convenient ``base-point''.



\subsection{Gluing construction of conformal blocks}

Useful
bases for the spaces of conformal blocks can be constructed by means of the gluing construction.
This construction allows one to construct conformal blocks on arbitrary Riemann surfaces $C$ from the 
conformal blocks associated to the three-punctured spheres appearing in a pants 
decomposition of $C$. For each of the simple closed curves used  
to define a given pants decomposition one has to specify a representation of the Virasoro algebra.
The gluing of the conformal blocks on the pairs of pants to a conformal block on $C$
is performed by summing over bases of the representations assigned to the cutting curves. 

To be specific, let us start from $C_{0,n}=\mathbb{P}^1\setminus\{z_1,\dots,z_n\}$ with 
a fixed pants decomposition defined by cutting $C_{0,n}$ along $n-3$ non-intersecting
simple closed curves $\ga_1,\dots,\ga_{n-3}$. Out of $C_{0,n}$ with the given
pants decomposition let us construct a $n+d$-punctured sphere
by first cutting $d$ sufficiently small non-intersecting discs $D_k$ around 
$z_k$, $k=1,\dots,d$, 
out of the pairs of pants appearing in the given pants decomposition for $C_{0,n}$.
Then glue twice-punctured discs $D_k'$ back in such a way that the resulting 
surface has punctures at $z_k$ and $u_k$, $k=1,\dots,d$. The boundary of $D_k'$ will be
denoted by $\mu_k$.
In the pair of pants containing $z_n$ let us finally replace a disc $D_{\infty}$ around $z_n$
by a three times punctured disc containing $z_n$, $z_0$ and $y$, with $z_0$ and $y$ contained
in a smaller disc $D_0$ inside of $D_{\infty}$. The boundaries of $D_{\infty}$
and $D_0$ will be denoted $\nu_\infty$ and $\nu_0$, respectively.
The result of this construction is a sphere with $n+d+2$ punctures $z_1,\dots,z_n;u_1,\dots,u_{d};z_0,y$
and a fixed pants decomposition.

The conformal blocks resulting from the gluing construction are determined uniquely up
to normalisation by the assignment of representations to the curves 
$\ga_r$ and  $\mu_k$ with $r=1,\dots,n-3$ and $k=1,\dots,d$, as well as 
$\nu_0$ and $\nu_\infty$.  
We will assign representations $V_{{Q}/{2}+\mathrm{i}p_k}$
to the curves $\ga_r$, $r=1,\dots,n-3$, and
representations $V_{\al_k}$ to the 
curves $\mu_k$, $k=1,\dots,d$. To the remaining curves 
$\nu_0$ and $\nu_\infty$ we will assign representations $V_0$ and  $V_{\al_n}$, 
respectively. 

The chiral partition functions of the conformal blocks defined in this way will be denoted 
$\CZ(\mathbf{p},\mathbf{u},\hat{\mathbf{z}};y)$ with  $\mathbf{p}=(p_1,\dots,p_{n-3})$.
$\CZ$ depends holomorphically on all of its variables.

\subsection{Classical limit of null vector decoupling equations} \label{S:BPZlim}

We will consider the limit $b\ra 0$ with $\de_r=b^2\De_r$ and $\la_r=bp_r$ kept fixed. 
One may notice that
the differential equations \rf{BPZ+} can be solved in the following form
\begin{equation}\label{factorclass}
\CZ(b^{-1}\mathbf{l},\mathbf{u},\hat{\mathbf{z}};y)=e^{-\frac{1}{b^2}\CW(\mathbf{l},\mathbf{u},\mathbf{z})}
\chi'(y|\mathbf{l},\mathbf{u},\mathbf{z})\chi''(z_0|\mathbf{l},\mathbf{u},\mathbf{z})(1+\CO(b^2)),
\end{equation}
where ${\mathbf{z}}=(z_1,\dots,z_{n-3})$ and $\mathbf{l}=(\la_1,\dots,\la_{n-3})$.
This ansatz will solve equation \rf{BPZlight} provided 
$\chi'$ and $\chi''$  are two solutions of the 
differential equation $(\pa^2_y+t(y))\chi(y)=0$, with  $t(y)$ of the form \rf{t-form},
where the parameters 
$v_r$ and $H_r$ are obtained from $\CW(\mathbf{l},\mathbf{u},\mathbf{z})$ as follows
\begin{equation}\label{genfct1}
v_r=-\frac{\pa}{\pa u_r}\CW(\mathbf{l},\mathbf{u},\mathbf{z})\,,
\qquad
H_r=\frac{\pa}{\pa z_r}\CW(\mathbf{l},\mathbf{u},\mathbf{z})\,.
\end{equation}
The  equations \rf{BPZheavy} furthermore imply that
the parameters $(u_r,v_r,H_r)$ are not independent but related through the constraints \rf{HJ}.
Equations \rf{projective} finally reproduce \rf{globalsl2}.

If $d=n-3$, one has just as many equations as one needs to determine $H_r$, $r=1,\dots,n$
as functions of $\mathbf{u},\mathbf{v}$, as was done in the definition of the Garnier system.
This is how the kinematics of the Garnier system is recovered from the classical limit of 
Virasoro conformal blocks in this case, as first observed in \cite{T10}. Similar observations have
been exploited in \cite{LLNZ}. The cases with $d<n-3$ will be discussed later.

\subsection{Verlinde loop operators} \label{S:Verlinde}

Useful additional information is provided by the action of the Verlinde loop operators
studied in \cite{AGGTV,DGOT} on spaces of conformal blocks, see \cite[Section 2.7]{T17a} 
for a brief review. 

The basic idea behind the definition of the Verlinde loop operators
is as follows. Given a conformal block $f$ on surface $C$, there is a canonical way to 
define a conformal block $f'$ on a surface $C'$ 
having an  extra puncture $y_0$
with vacuum representation assigned to $y_0$, and vice-versa.
A conformal block $f$  on  $C$  similarly defines a conformal block $f''$
on a surface $C''$ obtained by replacing  
a disc in $C$ by a disc $D_0$ containing two punctures at $y$ and $z_0$
with representations $V_{-b/2}$
assigned to both punctures, and vacuum representation assigned to the boundary of $D_0$.
If $f$ is a conformal block defined by the gluing construction one may use 
the null vector decoupling equations \rf{BPZlight}  to compute the analytic 
continuation $\ga.f''$of $f''$ along contours $\ga$ starting and ending at $y$.  
The contribution to $\ga.f''$ which has vacuum representation assigned to the boundary of $D_0$
may be canonically identified with a conformal block $\SL_\ga f$ on the original surface $C$. This defines
an operator $\SL_\ga$ on the space of conformal blocks associated to the surface $C$.
The algebra generated by the operators $\SL_\ga$ is  
a non-commutative deformation of the Poisson algebra of trace functions $L_\ga$ on $\CM_{\rm flat}(C)$
\cite{DGOT,TV13}.

We had previously associated  trace functions $L_{k,i}$, $i=s,t,u$ to 
each of the curves $\ga_k$ defining the pants decomposition of $C_{0,n}$.
The Verlinde loop operators associated to the contours defining $L_{k,i}$, $i=s,t,u$
will be denoted by $\SL_{k,i}$, for $i=s,t,u$ and $k=1,\dots,n-3$, respectively.

We will need the results of \cite{DGOT,AGGTV} for the Verlinde loop operators $\SL_{k,i}$
In the case of the operators $\SL_{k,s}$ a simple diagonal result was found,\footnote{Comparing 
to \cite{DGOT} we changed the definition of $\SL_{k,i}$ slightly to absorb a factor 
of $2\cos(\pi(1+b^2))$.}
\begin{subequations}\label{Verlinde}
\begin{equation}\label{Wilson}
\SL_{k,s}\CZ(\mathbf{p},\mathbf{u},\hat{\mathbf{z}};y)={2\cosh(2\pi bp_k)}
\CZ(\mathbf{p},\mathbf{u},\hat{\mathbf{z}};y).
\end{equation}
The expressions for the operators ${\SL}_{k,i}$ are  more complicated for $i=t,u$. They take the form
\begin{equation}\label{'tHooft}
{\SL}_{k,i}=\sum_{\nu=-1}^1 C^\nu_{k,i}(\mathbf{p})e^{\nu\,\mathrm{i}\, b\frac{\pa}{\pa{p_k}}}\,.
\end{equation}
\end{subequations}
Explicit formulae for the coefficients $C^\nu_{k,i}(\mathbf{p})$ can be found in \cite{AGGTV,DGOT}.

\subsection{Classical limit of Verlinde loop operators} \label{S:Verlinde}

Using the results of the explicit calculations of these
operators from \cite{AGGTV,DGOT} we will now identify the
variables $\la_k=bp_k$ and $\kappa_k=\frac{4\pi}{ {\mathrm{i}}}\pa_{\la_k}\CW$ with 
complex Fenchel-Nielsen coordinates on the character variety $\CM_{\rm char}(C_{0,n})$
in the limit $b\ra 0$ considered above.


In this limit one may 
compute the Verlinde loop operators in two different ways. One may, on the one hand, 
use the factorisation \rf{factorclass} in order to show that the classical limit of the
Verlinde loop operators can be identified with
traces of the monodromies of the differential operator $\pa^2_y+t(y)$.
This can be compared to the classical limit of the explicit formulae \rf{Verlinde} for the Verlinde
loop operators which turn out to be identical to the expressions for the trace functions given
in Section \ref{Darboux} when the conformal blocks are normalised appropriately.\footnote{Changing the normalisation of the three-point conformal blocks will change the form of the coefficients $C^\nu_{k,i}(\mathbf{p})$
in \rf{'tHooft}. There exists a choice of normalisation reproducing the corresponding coefficients in \rf{classWT}.}
In this way it
may be shown that $\CW(\mathbf{l},\mathbf{u},\mathbf{z})$ satisfies
\begin{equation}\label{Wtok}
\kappa_r=-4\pi\mathrm{i}\frac{\pa}{\pa \la_r}\CW(\mathbf{l},\mathbf{u},\mathbf{z}),\qquad r=1,\dots,n-3,
\end{equation}
with $\kappa_r=\kappa_r(\mathbf{l},\mathbf{u},\mathbf{z})$ being the value of the coordinate 
defined via \rf{classWT}.

\subsection{Classical limit of conformal blocks as generating function}

Recall that that the coordinates $(\mathbf{k},\mathbf{l})$ are a set of Darboux coordinates for the 
moduli space $\CM_{\rm flat}(C)$ of flat $\mathrm{SL}(2,\BC)$-connections on $C_{0,n}$.
Given the function $\CW(\mathbf{l},\mathbf{u},\mathbf{z})$
one may 
invert the relations \rf{Wtok} to define 
$\mathbf{u}=\mathbf{u}(\mathbf{k},\mathbf{l};\mathbf{z})$, and 
then define  $\mathbf{v}=\mathbf{v}(\mathbf{k},\mathbf{l};\mathbf{z})$ using \rf{genfct1}.
This is just the standard procedure to define a canonical transformation in 
Hamiltonian mechanics in terms of 
generating functions. The coordinates $(\mathbf{u},\mathbf{v})$ defined in this way will therefore be another
set of Darboux coordinates for the natural symplectic structure on $\CM_{\rm flat}(C)$ which is related to the  
natural symplectic structures of the
Garnier system as
\begin{equation}\label{AB-formrel}
\Omega=\frac{1}{\mathrm{i}}\sum_{r=1}^{n-3}dv_r\wedge du_r=\frac{1}{4\pi}
\sum_{r=1}^{n-3}d\kappa_r\wedge d\lambda_r\,.
\end{equation}

The Riemann-Hilbert correspondence defines a $\mathbf{z}$-dependent change of variables
from $(\mathbf{l},\mathbf{k})$ to $(\mathbf{u},\mathbf{v})$. Fixing 
$(\mathbf{l},\mathbf{k})$ by imposing the condition of
constant monodromy defines commuting flows of the variables 
$(\mathbf{u},\mathbf{v})\equiv (\mathbf{u}(\mathbf{z}),\mathbf{v}(\mathbf{z}))$. The 
Hamiltonian form \rf{Hameom} of the differential equations governing these flows
can be found as follows. We are considering  a 
canonical transformation in a non-autonomous
Hamiltonian system generated by the function $\CW$. Having dynamics in the variables 
$(\mathbf{u},\mathbf{v})$ described in Hamiltonian form \rf{Hameom}
is equivalent to having dynamics in the variables 
$(\mathbf{l},\mathbf{k})$ generated by the Hamiltonian $\tilde{H}_r=H_r-\pa_{z_r}\CW$.
For describing isomonodromic deformations we choose $\tilde{H}_r\equiv 0$, 
implying that the functions $H_r$ related to $\CW(\mathbf{l},\mathbf{u},\mathbf{z})$ via \rf{genfct1}
are the Hamiltonians to be used when representing the isomonodromic flows in the Hamiltonian form
\rf{Hameom}.  The fact that the functions $H_r$ defined from $\CW$ in \rf{genfct1} must coincide 
with the Hamiltonians of the Garnier system follows from the observation that both are uniquely 
determined by the system of linear equations \rf{HJ}.
We recover, in an independent way, the fact that the coordinates $(\mathbf{u},\mathbf{v})$
are  Darboux coordinates for the natural 
Poisson structure of the Garnier system.

In this way we have fully reproduced  the Hamiltonian representation of the 
Garnier system describing the isomonodromic deformations of the differential equation
$(\pa^2_y+t(y))\chi(y)=0$ with $t(y)$ of the form \rf{t-form} from conformal field theory. 

\section{Comparison with similar results}

We would here like to compare our results to some known results of a similar nature. 

\subsection{Genus zero analog of Kawai's theorem}\label{Kawai}

To begin with, let us discuss the cases where the number of degenerate fields $d$ is less than
$n-3$. The classical limit can be analysed in the same way as before, the only change being
a lower number of apparent singularities in the differential equation $(\pa^2_y+t(y))\psi(y)=0$.
In this case we can only determine a subset of the parameters $H_r$ using the 
constraints \rf{HJ}, determining, for example, $H_1,\dots,H_{d}$ as function
of the parameters $(\mathbf{u},\mathbf{v})$. The total number of independent variables
in the differential equation is $n-3+d$.

In the extreme case $d=0$ we do not have any apparent singularities, the only parameters
left are the $n-3$ independent variables $\mathbf{H}=(H_1,\dots,H_{n-3})$. These parameters can be 
identified as coordinates on the cotangent fibres of the moduli space $\CM_{0,n}$ of
$n$-punctured spheres  $C_{0,n}^\de$ having conical singularities at $z_r$, $r=1,\dots,n$ with deficit angles 
determined by the parameters $\de_r$ in \rf{t-form} \cite{TZ}.
Together with the positions 
$z_1,\dots,z_{n-3}$ 
one gets a system of Darboux 
coordinates for the total space of the cotangent bundle $T^*\CM_{0,n}$.

The holonomy of the corresponding connection defines a map from from $T^*\CM_{0,n}$ to the 
character variety $\CM_{\rm char}(C_{0,n})$. Parameterising 
points on $\CM_{\rm char}(C_{0,n})$ by the complex Fenchel-Nielsen coordinates 
introduced in Section \ref{Darboux} one obtains a change of coordinates 
from $(\mathbf{z},\mathbf{H})$ to $(\mathbf{k},\mathbf{l})$. 

In the  case $d=0$ presently under 
consideration we will still find a semiclassical asymptotics of the form 
\rf{factorclass}. 
However, the function $\CW$ characterising the leading term
will now be $\mathbf{u}$-independent, $\CW=\CW(\mathbf{l},\mathbf{z})$. 
This function will satisfy the relations \rf{Wtok} and the second relation in 
\rf{genfct1}, as before. These relations identify $\CW(\mathbf{l},\mathbf{z})$
as generating function for the change of coordinates 
from $(\mathbf{z},\mathbf{H})$ to $(\mathbf{k},\mathbf{l})$ in the sense of symplectic
geometry. The existence of such a generating function shows that the change of 
coordinates defined by the holonomy map preserves the natural symplectic structures. 
A similar result was obtained in
\cite{Ka} for the case of surfaces of type $C_{g,0}$.

Relations of the function $\CW(\mathbf{l},\mathbf{z})$ with conformal field theory
have been discussed in \cite{T10} and \cite{LLNZ}. It has been proposed in \cite{LLNZ}
to compute $\CW(\mathbf{l},\mathbf{z})$ from the limit of the isomonodromic deformation flows 
when $u_k$ approach $z_k$, $k=1,\dots,n-3$. 

The function $\CW(\mathbf{l},\mathbf{z})$ can furthermore be used 
to characterise the spectrum of the $\mathrm{SL}(2,\BC)$-Gaudin model \cite{T10,T17b}.
In the context of the Nekrasov-Shatashvili program relating supersymmetric gauge theories 
to integrable models
similar relations have been proposed
in \cite{NRS}.

\subsection{Relation to Liouville theory}

The observations made in Section \ref{Kawai} are related to the results obtained in \cite{TZ} describing the
Weil-Petersson symplectic form on the Teichm\"uller spaces $\CT_{0,n}$ in terms of 
the Liouville action functional. The metric of constant negative curvature $d^2s=e^{2\vf_u(y,\bar{y})} dyd\bar{y}$ 
on $C_{0,n}^\de$
defines a function $t_{u}(y)$ of the form \rf{t-form} with $d=0$ via $t_{u}(y)=-(\pa_y\vf_u)^2+\pa_y^2\vf_u$.  
The particular values of the residues $H_r$ that are found when $t(y)=t_u(y)$ will be denoted as
$E_r$.

The 
Liouville action is the functional of $\vf(y,\bar{y})$ having an extremum when $d^2s=e^{2\vf(y,\bar{y})} dyd\bar{y}$ 
has constant negative curvature. Evaluating the Liouville action at this extremum defines a function 
$S_{L}=S_{L}(\mathbf{z},\bar{\mathbf{z}})$.  The 
residues $E_r$ of $t_u(y)$ are related to $S_{L}$
as $E_r=\pa_{z_r}S_{L}(\mathbf{z},\bar{\mathbf{z}})$ \cite{TZ}.

One may note, on the other hand, that the holonomy of $t_{u}(y)=-(\pa_y\vf_u)^2+\pa_y^2\vf_u$ 
defines functions $(\mathbf{k}_u(\mathbf{z},\bar{\mathbf{z}}),\mathbf{l}_u(\mathbf{z},\bar{\mathbf{z}}))$. 
Based on the semiclassical limit of conformal field theory we will argue that
the function $\CW_u(\mathbf{z},\bar{\mathbf{z}})$
defined by restriction of $\CW(\mathbf{l},\mathbf{z})$ to 
$\mathbf{l}=\mathbf{l}_u(\mathbf{z},\bar{\mathbf{z}})$,
\begin{equation}\label{restr}
\CW_u(\mathbf{z},\bar{\mathbf{z}})=\CW(\mathbf{l}_u(\mathbf{z},\bar{\mathbf{z}}),\mathbf{z}),
\end{equation}
satisfies 
\begin{equation}\label{TZeqn}
\pa_{z_r}\mathrm{Re}(\CW_u(\mathbf{z},\bar{\mathbf{z}}))=H_r(\mathbf{z},\bar{\mathbf{z}}), \qquad r=1,\dots,n-3.
\end{equation}
Indeed, the correlation functions of Liouville theory, $\langle \prod_{r=1}^n e^{2\al_r\vf(z_r\bar{z}_r)}\rangle$,
can be decomposed into conformal blocks as \cite{ZZ,T01}
\begin{equation}\label{holofact} 
\bigg\langle \prod_{r=1}^n e^{2\al_r\vf(z_r\bar{z}_r)}\bigg\rangle=\int d\mu(\mathbf{p})\;\;
|\CZ(\mathbf{p},\mathbf{z})|^2.
\end{equation}
In the semiclassical limit one may use \rf{factorclass}. The integral over $\mathbf{p}=(p_1,\dots,p_{n-3})$ in \rf{holofact}
will be dominated by a saddle point determined by the condition that
\begin{equation}
\pa_{\la_k}\mathrm{Re}(\CW_u(\mathbf{l},\mathbf{z}))\Big|_{\mathbf{l}=\mathbf{l}_u(\mathbf{z},\bar{\mathbf{z}})}
=0, \qquad k=1,\dots,n-3.
\end{equation}
This condition ensures that $\CW_u(\mathbf{z},\bar{\mathbf{z}})$ 
defined in \rf{restr} satisfies \rf{TZeqn}. It follows that
the Liouville action $S_L$ coincides with $\mathrm{Re}(\CW_u)$ up to a constant. This 
observation is related to the 
characterisation of the spectrum of the $\mathrm{SL}(2,\BC)$-Gaudin model 
in terms of the function $\CW$ \cite{T10,T17b}.

\section{Conclusions: CFT as quantisation of the Garnier system}

The observations above relate conformal field theory to
the  quantisation of the  isomonodromic deformation problem. In order to quantise the Garnier system
one may start by observing that both
$(\mathbf{u},\mathbf{v})$ and $(\mathbf{l},\mathbf{k})$ represent Darboux coordinates 
for the moduli spaces of flat connections, \rf{AB-formrel}.
One may therefore consider two quantisation schemes in which  $u_r$ and $\lambda_r$ both get represented
as multiplication operators, whereas the operators associated to 
$v_r$ and $\kappa_r$ are $\sv_r=b^2\frac{\pa}{\pa u_r}$ and $\mathsf{l}_r=4\pi \frac{b^2}{\mathrm{i}}\frac{\pa}{\pa \la_r}$, 
acting on suitable  spaces of function $\Psi(\mathbf{u})$ and
$\Phi(\mathbf{l})$, respectively.

The next step will be to quantize the Hamiltonians $H_r$. We will define the quantum counterparts 
$\SH_r$ of $H_r$ as solutions to the following set of $n$ constraints,
\begin{subequations}
\begin{align}
&  b^2\frac{\pa^2}{\pa u^2_k}+\check{\mathsf{t}}_k(u_k) =0,\qquad
\check{\mathsf{t}}_k(u_k):=\lim_{y\ra u_k}\left(\mathsf{t}(y)+\frac{3+2b^2}{4(y-u_k)^2}-\frac{b^2}{y-u_k}
\frac{\pa}{\pa u_{k}}
\right), \\
&\mathsf{t}(y):=\sum_{r=1}^n
\left(\frac{b^2\De_r}{(y-z_r)^2}+\frac{\SH_r}{y-z_r}\right)-
\sum_{k=1}^{d} \left(\frac{3+2b^2}{4(y-u_k)^2}-\frac{b^2}{y-u_k}\frac{\pa}{\pa u_{k}}\right),\quad k=1,\dots,d,\notag \\
&\sum_{r=1}^nz^l_r\bigg(z_r\SH_r-(l+1)b^2\De_r\bigg)+
\sum_{k=1}^{n-3}u^l_k\bigg(b\,u_k\frac{\pa}{\pa u_{k}}-(l+1)b^2\De_{-\frac{1}{2b}}\bigg)=0,\qquad l=-1,0,1.
\notag\end{align}
\end{subequations}
As in the classical case one may solve these constraints to define second order differential operators
$\SH_r$ in the variables $\mathsf{u}$ if $d=n-3$. There are additional terms of order $b^2$ which ensure that
$[\SH_r,\SH_s]=0$ for all $r,s=1,\dots,n$.
It is then natural to require that the quantum Hamiltonians generate the 
evolution with respect to the ``time'' variables $z_r$,
\begin{equation}
b^2\frac{\pa}{\pa z_r}\Psi(\mathbf{u};\mathbf{z})=\SH_r\Psi(\mathbf{u};\mathbf{z})\,,\qquad r=1,\dots,n\,.
\end{equation}
These equations are easily seen to be equivalent to the null vector decoupling equations satisfied 
by the chiral partition functions $\CZ(\mathbf{p},\mathbf{u},{\mathbf{z}})$
of the conformal blocks on $C_{0,2n-3}$ obtained from the conformal blocks introduced in 
Section \ref{Set-up} by removing the punctures at $z_0$ and $y$, as was first pointed out in \cite{T10}.
Generalising \cite[Section 5.2]{GT} it is possible to show that 
these equation define the series expansion of $\CZ(\mathbf{p},\mathbf{u},{\mathbf{z}})$ associated
to suitable gluing patterns uniquely, with exponents of the 
leading terms in the expansions specified in terms of the variables $\mathsf{p}$.

It had previously been observed \cite{Re} that the Knizhnik-Zamolodchikov equations appearing 
in CFTs with affine Lie algebra symmetry can be interpreted as the time-dependent 
Schr\"odinger equations that would appear in the quantisation of the isomonodromic
deformation problem. Our observations in Section \ref{S:BPZlim} are close analogs of the 
results in \cite{Re} for the Virasoro case. Both are related to each other through a variant
of Sklyanin's Separation of Variables method \cite{T10}.
The consideration of the classical limit 
of the Verlinde loop operators in Section \ref{S:Verlinde} adds the crucial other side 
of the coin needed to get a precise characterisation of the classical conformal blocks
as generating functions.

The quantisation of this classical integrable system 
yields equations characterising the conformal blocks of the Virasoro algebra completely. 
Describing the conformal blocks  in this way suggests to 
reinterpret the 
chiral partition functions $\CZ(\mathbf{p},\mathbf{u},{\mathbf{z}})$ as the wave-functions 
intertwining the representations for the quantised Garner 
system in terms of functions $\Psi(\mathbf{u})$ and $\Phi(\mathbf{l})$
introduced above. 

We may furthermore note that
conformal field theory is related to the isomonodromic deformation
problem in two limits, the limit $c\ra\infty$ discussed here and the limit 
$c=1$ considererd in \cite{ILT}, see \cite{T17a} for a review.
In the case $c=1$ one may identify the isomonodromic 
tau-function with a Fourier-transformation of Virasoro conformal blocks.
This is remarkable, and deserves to be better understood.
 
{\bf Acknowledgements.} 
The author would like to thank N. Reshetikhin for interest in this work, 
and for the suggestion to include a comparison with similar results into 
the paper.

This work was supported by the Deutsche Forschungsgemeinschaft (DFG) through the 
collaborative Research Centre SFB 676 ``Particles, Strings and the Early Universe'', project 
A10.


\end{document}